\documentclass{article}

\usepackage{arxiv}
\usepackage[utf8]{inputenc} 
\usepackage[T1]{fontenc}    
\usepackage{lmodern}
\usepackage[colorlinks=true,allcolors=black,citecolor=red]{hyperref}       
\usepackage{url}            
\usepackage{booktabs}       
\usepackage{amsfonts}       
\usepackage{amsmath}
\usepackage{mathtools}
\usepackage{microtype}      
\usepackage{cite}
\usepackage{cancel}
\usepackage{subcaption}
\usepackage{colortbl}
\usepackage{svg}
\usepackage{pgf}
\usepackage{adjustbox}
\usepackage[linesnumbered,ruled,vlined]{algorithm2e}
\usepackage{epstopdf}

\newif\ifsubmission
\submissiontrue

\DeclareUnicodeCharacter{2212}{\textendash}

\usepackage{graphicx}
\usepackage{geometry}
\usepackage{color}
\usepackage{xcolor}

\DeclarePairedDelimiterX{\inp}[2]{\langle}{\rangle}{#1, #2}

\usepackage{bm}

\definecolor{olive}{cmyk}{0, 0, 1, 0.6}
\definecolor{darkblue}{rgb}{0, 0, 1}
\definecolor{darkred}{rgb}{.3, 0, 0}
\definecolor{darkgreen}{rgb}{0, 1, 0}
\definecolor{yellowgreen}{rgb}{0, 1, 0.7}
\definecolor{darkyellow}{rgb}{0.8, 0.7, 0.4}
\definecolor{plum}{rgb}{0.8,0, 1}
\definecolor{lightblue}{rgb}{0.7, 0.7, 1}
\definecolor{brown}{cmyk}{0, 0.8, 1, 0.6}
\definecolor{tangerine}{rgb}{1,0.65,0.2}
\definecolor{olive}{cmyk}{0, 0, 1, 0.6}
\definecolor{salmon}{rgb}{1,0.5,0.5}

\newcommand{\New}[1]{
    \ifsubmission
        {#1}
    \else
        \textcolor{darkred}{#1}
    \fi
}

\usepackage{setspace}
\usepackage{multicol}

\title{A Helmholtz equation solver using unsupervised learning: Application to transcranial ultrasound}

\author{
Antonio Stanziola$^{1,*}$, Simon R. Arridge$^{2}$, Ben T. Cox$^{1}$, and Bradley E. Treeby$^{1}$ \\
\\
$^{1}$Department of Medical Physics and Biomedical Engineering\\University College of London, Gower Street, London WC1E 6BT, UK\\ 
\\
$^{2}$Department of Computer Science\\University College of London, Gower Street, London WC1E 6BT, UK\\
\\
$^{*}$\texttt{a.stanziola@ucl.ac.uk}
}

\begin{document}
\maketitle
\begin{abstract}
Transcranial ultrasound therapy is increasingly used for the non-invasive treatment of brain disorders. However, conventional numerical wave solvers are currently too computationally expensive to be used online during treatments to predict the acoustic field passing through the skull (e.g., to account for subject-specific dose and targeting variations). As a step towards real-time predictions, in the current work, a fast iterative solver for the heterogeneous Helmholtz equation in 2D is developed using a fully-learned optimizer. The lightweight network architecture is based on a modified UNet that includes a learned hidden state. The network is trained using a physics-based loss function and a set of idealized sound speed distributions with fully unsupervised training (no knowledge of the true solution is required). The learned optimizer shows excellent performance on the test set, and is capable of generalization well outside the training examples, including to much larger computational domains, and more complex source and sound speed distributions, for example, those derived from x-ray computed tomography images of the skull. 
\end{abstract}

\keywords{Helmholtz equation \and learned optimizer \and unsupervised learning \and physics-based loss function \and transcranial ultrasound}

\section{Introduction and background}


\subsection{Motivation}

Transcranial ultrasound therapy is a rapidly emerging technique for the noninvasive treatment of brain disorders in which ultrasound is used to cause functional or structural changes to brain tissue. Several different types of treatment are possible depending on the pattern of ultrasound pulses used and the addition of exogeneous microbubbles. This includes precisely destroying small regions of tissue \cite{mcdannold2010transcranial}, generating or suppressing electrical signals in the brain \cite{legon2014transcranial}, and temporarily opening the blood-brain barrier to allow drugs to be delivered more effectively \cite{abrahao2019first}. A major challenge for transcranial ultrasound therapies is the presence of the skull bone, which causes the ultrasound waves to be distorted and attenuated, even at low frequencies \cite{hynynen1998demonstration}. Critically, the skull morphology and acoustic properties vary both within and between individuals, which leads to undesirable changes in the position and intensity of the ultrasound focus \cite{chang2016factors,gimeno2019experimental}, and in some cases can destroy the focus entirely \cite{sun1998focusing}.

Using computational ultrasound models and knowledge of the geometric and acoustic properties of the skull (e.g., derived from an x-ray computed tomography image), it is possible to predict the ultrasound field inside the brain after propagating through the skull, and thus account for subject-specific dose and targeting variations \cite{aubry2003experimental,lee2016transcranial}. However, existing models based on conventional numerical techniques typically take tens of minutes to several hours to complete due to the large size of the computational domain compared to the size of the acoustic wavelength, in some cases generating models with billions of unknowns which require tens of thousands of iterations to solve  \cite{pinton2011effects,pulkkinen2014numerical,almquist2016rapid,robertson2017accurate,mcdannold2019elementwise}. This makes them too slow to be used for online calculations and corrections, i.e., while the subject is undergoing the therapy. Consequently, approximate models are often used (e.g., ray tracing) which trade off between accuracy and computational efficiency \cite{clement2002non,kyriakou2015full}.

In the current work, instead of using a classical numerical partial differential equation (PDE) solver, a recurrent neural network architecture is developed to rapidly solve the Helmholtz equation with a heterogeneous sound speed distribution representative of a human skull. The network is trained using a physics loss term based on the Helmholtz equation and a training set of idealized sound speed distributions. The use of a physics loss term, which plays an analogous role to the data consistency term in inverse problems \cite{JinUnser2017,hammernik2019sigma}, avoids the need to run a large number of computationally expensive simulations using a conventional solver to generate training data for supervised training. A review of the relevant background to this approach is described in the remainder of \S1, with the developed network architecture and training outlined in \S2. Results are then given in \S3 with discussion and outlook in \S4.

\subsection{Governing equations}

In the most general case, the propagation of ultrasound waves through the skull and brain involves a heterogeneous distribution of material parameters, shear wave effects in the skull, nonlinear effects when high intensities are used, and acoustic absorption \cite{white2006longitudinal,pulkkinen2014numerical}. However, if the ultrasound waves approach the skull close to normal incidence (this is often the case), then shear motion can be ignored \cite{robertson2017sensitivity}. Moreover, nonlinear effects are only important for ablative therapies and are restricted to a small region near the focus \cite{rosnitskiy2019simulation}, and acoustic absorption can be considered a second-order effect \cite{treeby2012modeling}. In addition, for many therapeutic applications, the applied ultrasound signals are at a single frequency and last for many milliseconds or seconds, which is typically much longer than the time taken for the acoustic field to reach a steady-state, and thus time-independent models can be used.

\begin{figure}[!t]
    \centering{
    \def\svgwidth{0.3\linewidth}
\begingroup%
  \makeatletter%
  \providecommand\color[2][]{%
    \errmessage{(Inkscape) Color is used for the text in Inkscape, but the package 'color.sty' is not loaded}%
    \renewcommand\color[2][]{}%
  }%
  \providecommand\transparent[1]{%
    \errmessage{(Inkscape) Transparency is used (non-zero) for the text in Inkscape, but the package 'transparent.sty' is not loaded}%
    \renewcommand\transparent[1]{}%
  }%
  \providecommand\rotatebox[2]{#2}%
  \newcommand*\fsize{\dimexpr\f@size pt\relax}%
  \newcommand*\lineheight[1]{\fontsize{\fsize}{#1\fsize}\selectfont}%
  \ifx\svgwidth\undefined%
    \setlength{\unitlength}{175.35211654bp}%
    \ifx\svgscale\undefined%
      \relax%
    \else%
      \setlength{\unitlength}{\unitlength * \real{\svgscale}}%
    \fi%
  \else%
    \setlength{\unitlength}{\svgwidth}%
  \fi%
  \global\let\svgwidth\undefined%
  \global\let\svgscale\undefined%
  \makeatother%
  \begin{picture}(1,0.93870298)%
    \lineheight{1}%
    \setlength\tabcolsep{0pt}%
    \put(0.45,0.38){\color[rgb]{0,0,0}\makebox(0,0)[lt]{\lineheight{1.25}\smash{\begin{tabular}[t]{l}$c(r)$\end{tabular}}}}%
    \put(0,0){\includegraphics[width=\unitlength,page=1]{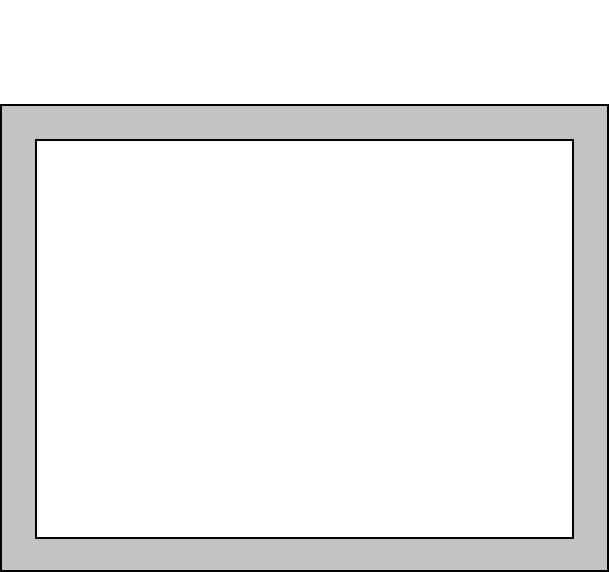}}%
    \put(0.12764939,0.5684062){\color[rgb]{0,0,0}\makebox(0,0)[lt]{\lineheight{1.25}\smash{\begin{tabular}[t]{l}$\Omega$\end{tabular}}}}%
    \put(0,0){\includegraphics[width=\unitlength,page=2]{images/domain.pdf}}%
    \put(0.15179803,0.88604113){\color[rgb]{0,0,0}\makebox(0,0)[lt]{\lineheight{1.25}\smash{\begin{tabular}[t]{l}$\textnormal{PML}$\end{tabular}}}}%
    \put(0,0){\includegraphics[width=\unitlength,page=3]{images/domain.pdf}}%
  \end{picture}%
\endgroup%

    \caption{Definition of the computational domain $\Omega$ which contains a heterogeneous sound speed $c(r)$, in this case represented by a skull. A perfectly matched layer (PML) is used to surround the computational domain to simulate exterior open boundaries as discussed in Appendix \ref{sec:appendix}.}
    \label{fig:domain_definition}
    }
\end{figure}

Considering the above, a simplified model of wave propagation through the skull and brain can be described by the heterogeneous Helmholtz equation subject to the Sommerfeld radiation condition at infinity:
\begin{align}
   \left[\nabla^2 + \left(\frac{\omega}{c(r)}\right)^2 \right]u(r) &= \rho(r) \enspace, \label{eq:helmholtz}\\
   \textnormal{s.t.} \; \lim_{|r| \to \infty} |r|^{\frac{n-1}{2}} \left( \frac{\partial}{\partial |r|} - i\frac{\omega}{c_0} \right) u(r) &= 0 \enspace. \label{eq:sommerfeld}
\end{align}
Here $n$ is the number of spatial dimensions, $c:\mathbb{R}^n\to \mathbb{R}^+$ is the speed of sound, $\omega$ is the angular frequency of the source, $r\in \mathbb{R}^n$ is a general space coordinate, $\rho:\mathbb{R}^n\to \mathbb{C}$ is the source distribution, 
and $u(r)\in \mathbb{C}$ is the complex acoustic wavefield. 
Here, it is assumed that the speed of sound distribution $c(r)$ is heterogeneous in a bounded region of the domain, while it is uniform and equal to $c_0$ outside of it.
In practice, the solution to Eq.~\eqref{eq:helmholtz} is sought within a domain of interest $\Omega \subset \mathbb{R}^n$ as shown in Fig.\ \eqref{fig:domain_definition}.

The aim of the current work is to find a learned iterative solver capable of generating a field $u$ which satisfies Eqs.\ \eqref{eq:helmholtz} and \eqref{eq:sommerfeld}. We will focus on the 2D version of the Helmholtz problem, keeping in mind that the ultimate goal is to translate our findings to large three-dimensional simulations. Extensions to 3D and to other wave equations, e.g., to include the effects of nonlinearity, changes in mass density, and acoustic absorption, will be considered as part of future work.


\subsection{Numerical solution techniques}
\label{sec:numerical_solution}

The Helmholtz equation given in Eq.\ \eqref{eq:helmholtz} can be written in the general form
\begin{equation}
    A(c)u = \rho \enspace,
    \label{eq:linear_operator}
\end{equation}
where $A(c)$ is a linear forward operator which depends on the speed of sound distribution $c$. There are many approaches to discretize $A$, including finite difference methods \cite{wang2010acoustic}, boundary element methods \cite{gumerov2009broadband}, and finite-element methods \cite{bermudez_optimal_2007}. In many cases, direct inversion of the forward operator is not feasible due to the size and conditioning of the problem, and thus iterative schemes are used. In this case, the solution of the PDE is cast as an optimization problem via a suitable loss function, which is solved using a minimization algorithm. 

The most widely used loss $L$ is the squared norm of the residual $e_k$, which is equivalent to the mean squared error (MSE) up to a scaling factor. Given the solution $u_k$ at iteration $k$ over the domain of interest $\Omega$, this can be calculated by \cite{raissi_physics_2017}:
\begin{equation}
    L_k(u_k,c,\rho) =  \|e_k \|^2 = \int_\Omega |e_k |^2 d r \enspace, \qquad e_k = A(c)u_k - \rho \enspace.
    \label{eq:mse_physics_loss}
\end{equation}
Other objective functions may also be used depending on the required characteristics of the solution or the discretization model at hand. Common alternative choices are the root-MSE (RMSE) or the mean absolute error (MAE). Other examples relevant to the solution of PDEs include using the Dirichlet energy  \cite{e_deep_2018}, augmenting the MSE loss with an $\ell_\infty$ term to enforce strong PDE solutions \cite{bar_unsupervised_2019}, and using physics-constrained loss functions \cite{zhu_physics-constrained_2019}.

Given a suitable discretization and loss function, a common approach to solving the system of equations is the use of Krylov subspace methods, such as the widely used conjugate gradient (CG) and generalized minimal residual (GMRES) methods \cite{saad_gmres_1986}. However, Krylov subspace methods are known to have a slow convergence rate for the Helmholtz problem \cite{ernst2012difficult}. From an intuitive perspective, if the solution starts with an empty wavefield and a spatially localized source, the nature of the Helmholtz equation makes each update of Krylov methods local (due to the Laplacian).\footnote{Krylov methods work by building a sequence of basis vectors $v_{j+1} = A^{\rm T}Av_j$ and finding the optimal solution in the subspace spanned by this sequence up to the current value of $j$. Thus the spatial support of the solution can only increase by the support of $A^{\rm T}A$ on each iteration, which is local for differential operators.} This means the solution will grow slowly from the source position (an example is shown later in the results section in Fig.~\ref{fig:evolution_of_solution}). However, the solution to the heterogeneous Helmholtz problem clearly has non-local dependencies, for example, a strong reflector at one end of the domain will affect the wavefield at the other end.

To mitigate these issues, preconditioning methods can be used. Mathematically, this means finding a suitable change of basis that reduces the dynamic range of the singular values of the forward operator, which often leads to the use of a multiscale representation that can take into account long-range dependencies \cite{erlangga_novel_2006, koren_shifted-laplacian_2009}. However, finding suitable preconditioning methods for wave problems is a challenging task. 
\New{Algebraic methods, such as sweeping preconditioners, are often based on matrix factorization methods that require instantiating the numerical discretization of the operator \cite{engquist2011sweeping}, which is often prohibitive for spectral methods that implicitly assume dense matrices.}

\New{Standard approaches to adapt multigrid preconditioners for the Helmoltz equation require careful rethinking of the standard components of multigrid methods, especially for coarse grids. This makes off the shelf multigrid methods, such as those based on damped Jacobi smoothing, effective only for low wavenumbers \cite{elman2001multigrid}. 
One of the first successful attempts is given in  \cite{elman2001multigrid}, where the authors used GMRES as a smoothing operator to effectively suppress high-frequency components. While this produced impressive convergence results for large-wavenumber problems, each iteration is computationally demanding due to the internal iterative solvers applied before each restriction operation. This idea was extended in \cite{chen2020meta} by learning the optimal subspace for GMRES, instead of relying on the classical Krylov subspace, although results are only shown for parabolic PDEs.}

\New{Other authors proposed  to use a shifted-Laplacian preconditioner which can be inverted using multigrid methods \cite{erlangga_novel_2006}. While effective, the multigrid iteration used for preconditioner inversion is still limited in its deepest coarse correction by the eigenvalue structure of the preconditioner itself, which is close to the original Helmholtz operator. Furthermore, convergence analysis and computation of the preconditioner's inverse is in most cases performed with a finite-difference discretization \cite{koren_shifted-laplacian_2009}, often relying on the sparse structure of the discretized operator to perform efficient computations, which is not applicable to spectral discretizations that produce much denser matrices. The lack of a sparse structure is also a challenge for incomplete LU decomposition techniques.}

An alternative idea towards improving Krylov-based iterative schemes is given in  \cite{rizzuti_learned_2019}, where Krylov iterations are interleaved with a UNet \cite{ronneberger2015u} that \textit{boosts} the current solution towards the true one. In \cite{rizzuti_learned_2019}, the UNet is trained on known examples, however, training on unlabeled examples using a physics loss is also suggested. A nice advantage of keeping some Krylov iterations is that they may act as regularizers, preventing the solution from diverging.

More generally, this leads to the following question: instead of using traditional optimizers to solve Eq.\ \eqref{eq:linear_operator}, is it possible to \textit{learn} a suitable optimizer $f$ that can be iteratively applied to generate an accurate solution in a small number of iterations, while at the same time being simple enough to be used on large scale problems? While the general idea of learning an optimizer is not new \cite{andrychowicz_learning_2016,putzky_recurrent_2017}, fully learned optimizers for the Helmholtz equation have not previously been explored, and the task involves a problem-specific trade-off between speed (number of iterations) and accuracy (how close we are to the solution).


\subsection{Learned optimizers}

In the case of the heterogenous Helmholtz equation given in Eq.\ \eqref{eq:helmholtz}, assuming a fixed angular frequency $\omega$, an iterative optimization scheme could be written in the form
\begin{equation}
    u_{k+1} = u_k + f_\theta(u_k, c, \rho),
    \label{eq:iterative_update_helmholz}
\end{equation}
where $f_\theta$ is a learned function parametrised by the estimate of the wavefield $u_k$ after the $k^{\mathrm{th}}$ iteration, the sound speed distribution $c$ (which is used in the forward operator $A$), and the source distribution $\rho$. However, a key problem with this formulation is that it is hard for the network to directly manipulate the sound speed and source distributions to give an update for $u$, as they belong to two very different domains. Although some learning-based methods have been proposed to combine samples from different domains, such as AUTOMAP \cite{zhu_image_2018}, it is hard to design a function approximator with the right inductive biases while preserving the necessary flexibility required for fast inference. This often means that fully connected deep networks must be used, which are hard to scale to high dimensional input/output pairs and require a large amount of data to be trained. 

Instead of using Eq.\ \eqref{eq:iterative_update_helmholz}, we can instead leverage our knowledge of the forward operator to manipulate $c$ and $\rho$ to get a new input $e_k$, which belongs to the same domain as $u_k$. For example, we could use the residual signal $e_k$ as an input
\begin{equation}
    u_{k+1} = u_k + f_\theta(u_k,e_k) \enspace, \qquad e_k = A(c)u_k - \rho \enspace.
    \label{eq:update_with_residual}
\end{equation}
(Other choices are also possible for $e_k$, for example, the derivative of the loss function.) This approach is in direct parallel to the inputs to many optimization algorithms. However, Eq.\ \eqref{eq:update_with_residual} still assumes that the combination of the current wavefield estimate and latest residual gives enough information to the network to specify the next update. This is unlikely to be true, as for different problems, the same residual could be observed given the same wavefield. A simple example is for an empty wavefield and a fixed source distribution $\rho$, the residual will be the same regardless of the sound speed distribution. If we look at the choice of the sequence of updates as a Markov decision process (MDP) \cite{li_learning_2016}, this makes it only partially observable. 

One way to restore the properties of a fully observable MDP is to construct a recurrent \textit{belief state} $h$ which contains information from all preceding actions and observations \cite{hausknecht_deep_2017}. Augmenting the problem with such a state can be done via the following update rule:
\begin{align}
    (\Delta u_{k+1}, h_{k+1}) &=  f_\theta(u_k,e_k,h_k) \nonumber \\
    u_{k+1} &= u_k + \Delta u_{k+1} \enspace.
    \label{eq:recurrent_update}
\end{align}
Note that the wavefield update resembles a discrete Euler stepping scheme \cite{lu_beyond_2020}. Augmenting the input with a hidden state is known to improve the representation capabilities of neural ODEs \cite{dupont_augmented_2019} and it is therefore reasonable to assume it also helps in their discretized counterpart. 

If the function $f_\theta$ is considered as an iterative solver, the presence of the state variable $h$ allows several optimizers to be cast in this framework. For example, if $h$ stores the previous gradient and its magnitude, $f_\theta$ could in principle work as a quasi-Newton method \cite{andrychowicz_learning_2016,putzky_recurrent_2017}, while if it stores the collection of all the previous residuals, it may generalize Krylov subspace methods.

A learned iterative solver in this framework was proposed by Putzky and Welling \cite{putzky_recurrent_2017} for various kinds of image restoration tasks in which $e_k$ was given by the gradient of the loss function (in their case, the log likelihood for a Gaussian prior). This scheme was later extended by Adler and Oktem \cite{adler_solving_2017} for non-linear operators and applied to a non-linear tomographic inversion problem.
\section{Network architecture and training}


\subsection{Iterative solution to the Helmholtz equation}

\begin{figure}[tp]
    \centering
    \ifsubmission
        \includegraphics[width=\textwidth]{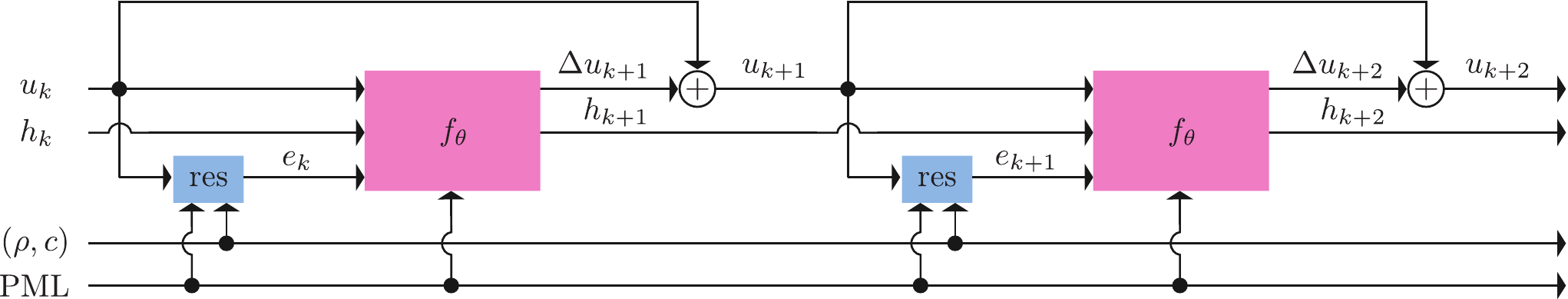}
    \else
        \includegraphics[width=\textwidth]{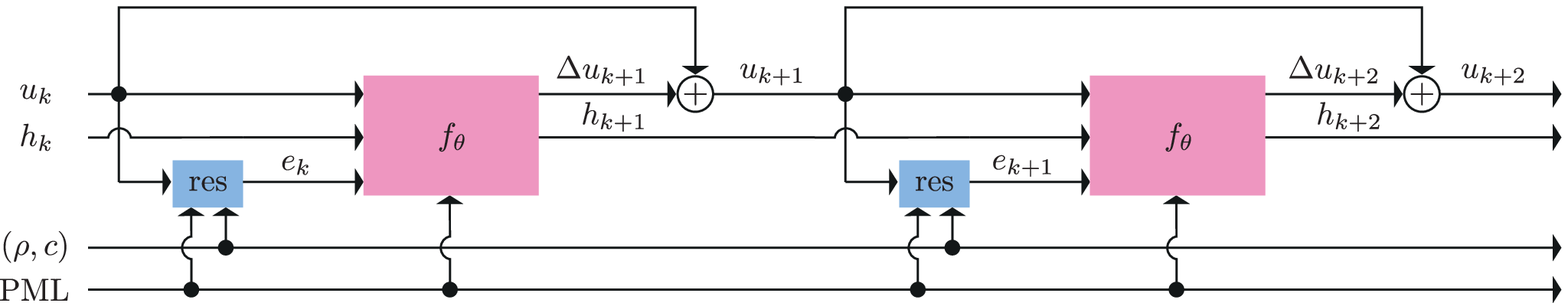}
    \fi
    \caption{Schematic of the proposed iterative scheme (showing two iterations) for the solution of the heterogeneous Helmholtz equation $u$. This uses a fully-learned optimizer $f_\theta$ which outputs both the solution update $\Delta u_{k}$ along with a belief state $h_{k}$. The residual $e_k$ is calculated using a physics loss (represented in the figure by the block $\mathrm{res}$). This is different for every speed of sound distribution and is used as an additional input to the network. Both the optimizer and the residual calculation make use of knowledge of the perfectly matched layer (PML). The residual calculation also uses the sound speed $c$ and source distribution $\rho$ through the forward operator.}\label{fig:unrolled_iteration}
\end{figure}

Building on the work discussed in \S1.4, we propose an iterative method in the form of Eq.\ \eqref{eq:recurrent_update} to solve the Helmholtz equation using a learned optimizer. The discrete dynamical system that models the iterative solution $u_k$ of the heterogeneous Helmholtz equation given in Eq.\ \eqref{eq:helmholtz} can be written as
\begin{align}
    e_k &= \left[\nabla^2 + \left(\frac{\omega}{c}\right)^2 \right]u_k - \rho\nonumber\\
    (\Delta u_{k+1}, h_{k+1}) &= f_\theta(u_k, e_k, h_k) \nonumber \\
    u_{k+1} &= u_k + \Delta u_{k+1} \enspace.
    \label{eq:iterative_solver}
\end{align}
Here $f_\theta$ is the neural network with learnable parameters $\theta$, $e_k$ is the residual computed from the heterogeneous Helmholtz equation, $\Delta u_k$ is the iterative update, and $h_k$ is a learned hidden or belief state. A diagram showing two unrolled iterations is given in Fig.\ \ref{fig:unrolled_iteration}. The network also uses an auxiliary input of the variable absorption coefficients that characterize the perfectly matched layer (PML) in each Cartesian direction (the PML is used as part of the discretized Laplacian operator to mimic the radiation condition in Eq.~\eqref{eq:sommerfeld} as outlined in Appendix \ref{sec:appendix}). This allows the network to learn how to dampen the waves at the edges of the domain to simulate exterior open boundaries, and to appropriately weight the corresponding residuals.

\begin{figure}[tp]
    \centering
    \ifsubmission
        \includegraphics[width=\textwidth]{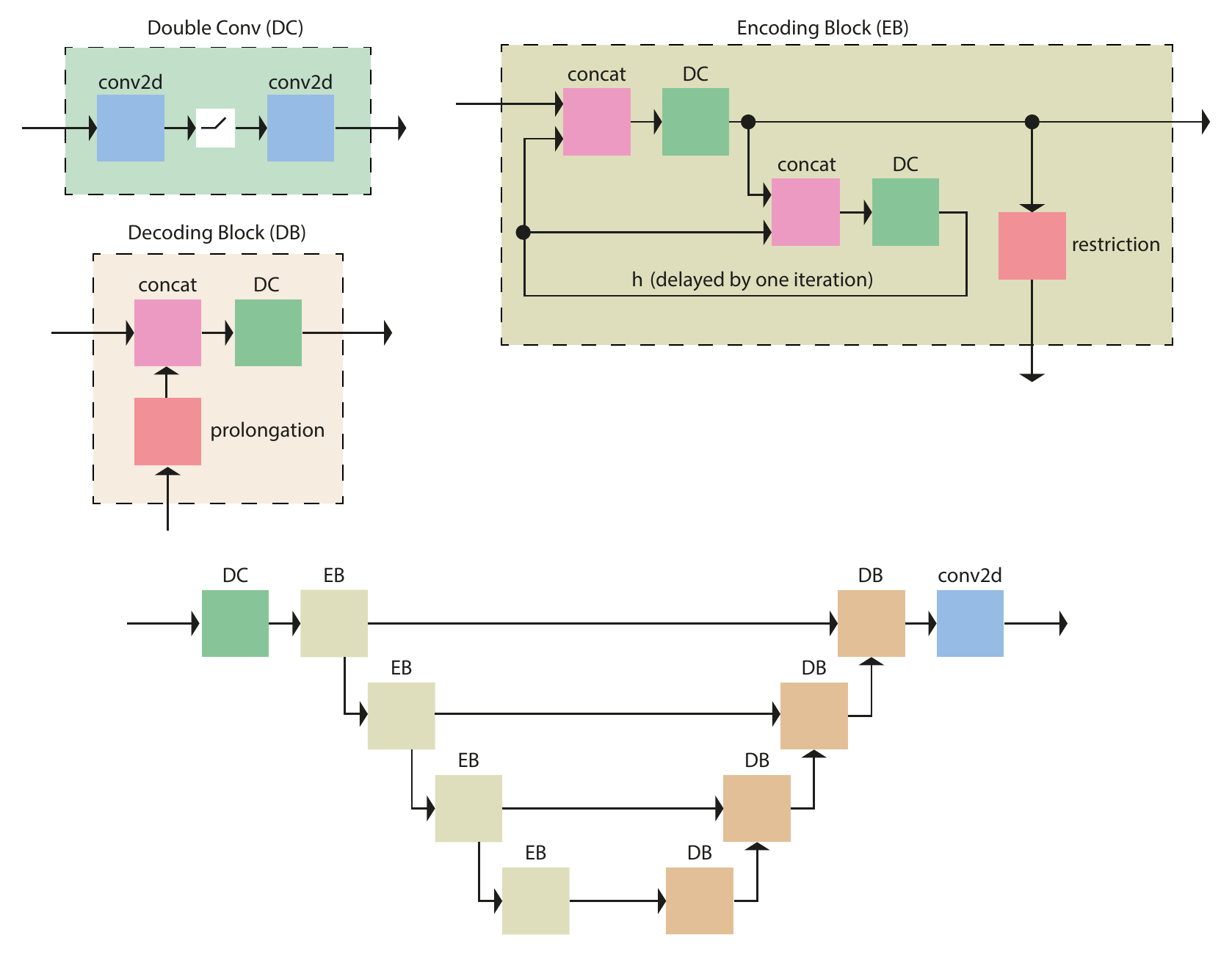}
    \else
        \includegraphics[width=\textwidth]{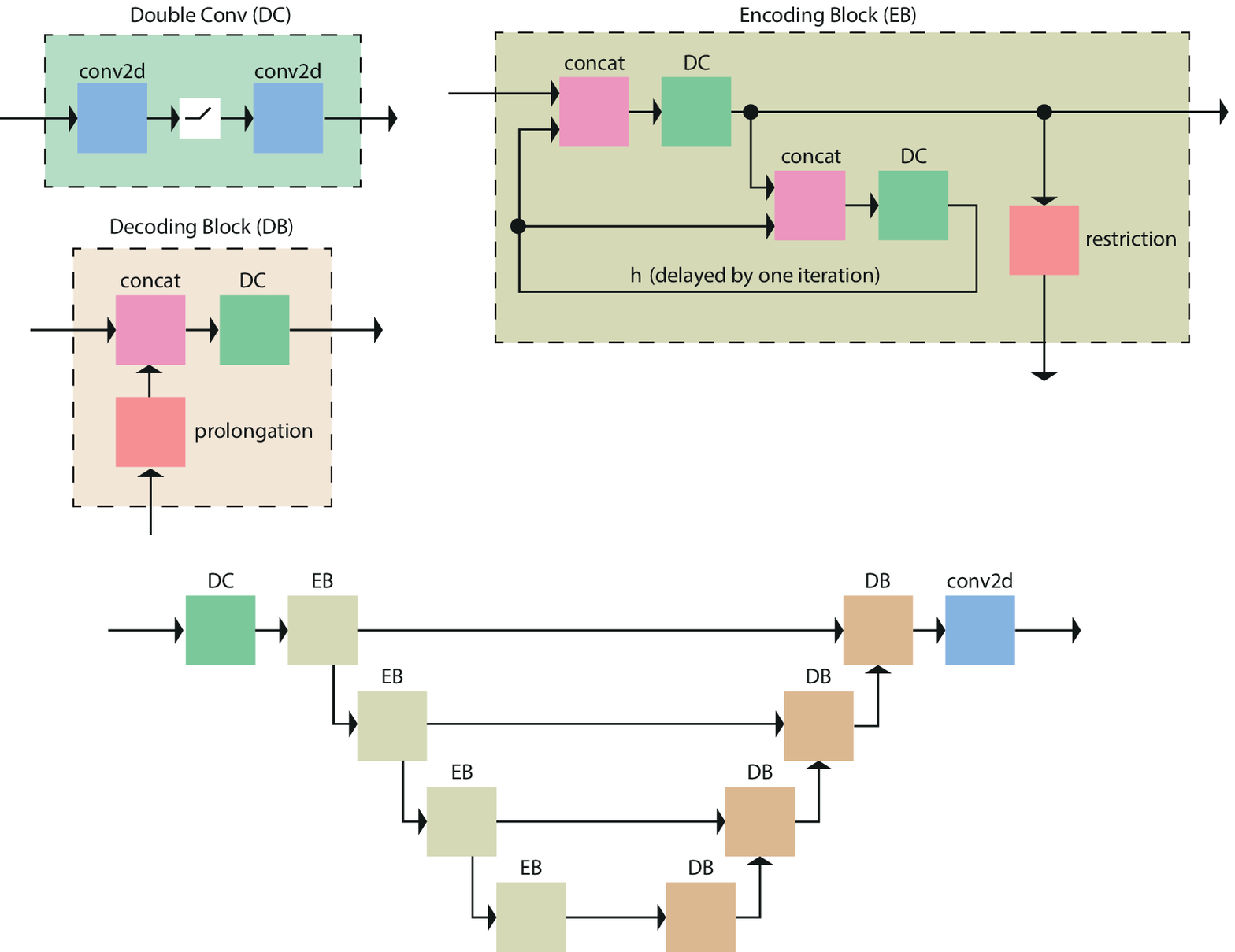}
    \fi
    \caption{Architecture of the modified UNet used for the learned optimizer $f_\theta$. Each encoding block (EB) contains two double convolution (DC) layers, one to compute the output passed to subsequent layers, and one to compute the hidden state $h$. The concat blocks stack the inputs in the channel dimension.   The network is lightweight, with only 8 channels per convolutional block at every scale and a total of 47k trainable parameters.}
    \label{fig:nn_architecture}
\end{figure}

The network is trained using a physics-based loss function. This avoids the need for labeled training data, e.g.\ created by running a large number of simulations using a conventional PDE solver to obtain ground truth solutions. While generating sufficient labeled training data may technically be feasible in 2D, for 3D problems, running a single simulation can take tens of minutes to many hours \cite{pulkkinen2014numerical}, which makes the generation of a large training set practically intractable. 

If the total number of iterations is fixed to $T$ (i.e., a finite horizon), one approach for training the network would be to minimize $L_{T}$, the final loss calculated after performing $T$ iterations. However, in practice,  it is not computationally feasible to perform backpropagation through a large number of iterative steps of the learned optimizer. The optimal choice of $T$ is also not straightforward, as the accuracy required and the number of iterations to reach that accuracy are problem specific. 

Instead, we choose to minimize the total loss function $R$ across all iterations, which is given by
\begin{equation}
    R = \sum_{k=0}^{T} w_k L_k(u_k, c, \rho) \enspace,
    \label{eq:weighted_iterative_loss}
\end{equation}
where $L_k$ is the physics-based loss function calculated after each iteration, and $w_k \geq 0$ are a set of weights that define the importance of the loss at each step. Here, the loss from Eq.\ \eqref{eq:mse_physics_loss} is used with the residual from Eq.\ \eqref{eq:iterative_solver}. The weights are set to $w_k = 1 \, \forall \, k$, in other words, we aim to minimize the loss at every step of the iteration, on average, which yields 

\begin{equation}
    R = \sum_{k=0}^{T} \|e_k \|^2 \enspace.
    \label{eq:linear_quadratic_tracking}
\end{equation}

To avoid unrolling the network for a large number of iterations for backpropagation, a replay buffer is used as discussed in detail in \S \ref{sec:training}.


\subsection{Neural network architecture}

The neural network architecture for the optimizer $f_\theta$ is depicted in Fig.\ \ref{fig:nn_architecture}. The input to the network has the same spatial  dimensions as the sound speed distribution (which is defined on a regularly spaced Cartesian grid) and contains six channels: two channels for the real and imaginary parts of the wavefield $u_k$, two channels for the real and imaginary parts of the residual $e_k$, and two channels for the variable absorption coefficients used in the PML in each Cartesian direction ($\sigma(x)$ and $\sigma(y)$ defined in Appendix \ref{sec:appendix}). 

The core building block of the network is the widely used double-convolution layer, which consists of two bidimensional convolutions with 3-by-3 kernels, interleaved by a parametrized-ReLU non-linear activation function \cite{he2015delving}. Each encoding block contains two double-convolution layers. The first accepts two inputs: the network input representation at the current scale and a hidden state. The output is then passed to a second double-convolution layer (used to update the hidden state), the corresponding decoding block of the network, and a restriction operator which downsamples the output and feeds it to a deeper layer of the network. The restriction operator is represented by an 8-by-8 convolutional stage, applied with stride 2 in order to halve the dimension of the wavefield at each depth, reaching a total of 4 depths.

Internally, each encoding block stores its own hidden state, which from a functional point of view can be considered as being passed between iterations, as shown in Fig.~\ref{fig:unrolled_iteration}. The size of the hidden state for each encoding block matches the size of its corresponding input, and has two channels. Note, the state variable $h_k$ in Eq.\ \eqref{eq:iterative_solver} and Fig.~\ref{fig:unrolled_iteration} refers to all the states stored across all encoding blocks.

The decoding blocks take an input from the layer below, upsamples it using transposed convolutions with 8-by-8 kernels and stride 2, and after concatenating it with the output from the corresponding encoding block (i.e., the skip connection), produces an output via another double-convolution layer. Finally, the last layer of the network is a 1-by-1 convolution that maps the output of the neural network to the wavefield domain. The output has the same spatial dimensions as the input and contains two channels for the real and imaginary parts of the wavefield.

There are several intuitions behind this choice of architecture. First, having a fully convolutional network implicitly imposes some degree of translation invariance to the iterative solver,\footnote{For example, if $\rho$ and $c$ were shifted upwards by an equal amount, we would expect the solution also to be shifted upwards by the same amount, i.e., for it to be translationally invariant. Therefore it is desirable for the neural network architecture to also be translationally invariant.}  while at the same time allows the network to be used with arbitrarily-sized sound speed distributions. Second, the network can encode priors at different scales thanks to the multiscale structure, allowing correction for very local distortions of the wavefield while at the same time taking care of long range dependencies. This is very similar to the idea behind the MultiGrid Neural Network (MgNet) \cite{he_mgnet_2019}, which connects UNet-like architectures with the theory of multiscale solvers, the latter widely used to solve the Helmholtz problem.

In total, the network has approximately 47k trainable parameters, with only 8 channels per convolutional block at every scale. The very small network size is possible because the solution is iteratively updated using the residuals of the true forward operator. \footnote{Code is publically available at \url{https://github.com/ucl-bug/helmnet}
}


\subsection{Training}
\label{sec:training}

The neural network is trained on a dataset of sound speed distributions containing idealized skulls. All calculations are performed in normalized units with a source frequency of $\omega = 1$ rad/s and a background sound speed of 1 m/s. The idealized skulls are randomly generated  with a hollow convex structure with a constant thickness and constant speed of sound, defined by summing up several circular harmonics of random amplitude and phase. Between examples, the skull thickness ranges from 2 to 10 m and the sound speed from 1.5 to 2 m/s giving a maximum sound speed contrast of 100 \% (this matches the sound speed contrast between soft tissue and human skull bone \cite{fry1978acoustical}). The size of each example is 96 $\times$ 96 grid points with a normalized grid spacing of 1 m, giving $2\pi$ points per acoustic wavelength (PPW). Note that while the training is performed using normalized units, the results can be re-scaled for any combination of grid spacing / source frequency / background sound speed, provided the PPW remains fixed at $2\pi$ (an example of simulating a transcranial ultrasound field within an adult skull at 490 kHz is discussed in Sec.\ \ref{sec:results_generalizability}). 

\begin{figure}[t]
    \centering
    \begin{adjustbox}{clip,trim=0cm 1.6cm 0cm 1.6cm}
    \makebox[\linewidth]{
        \scalebox{0.6}{\input{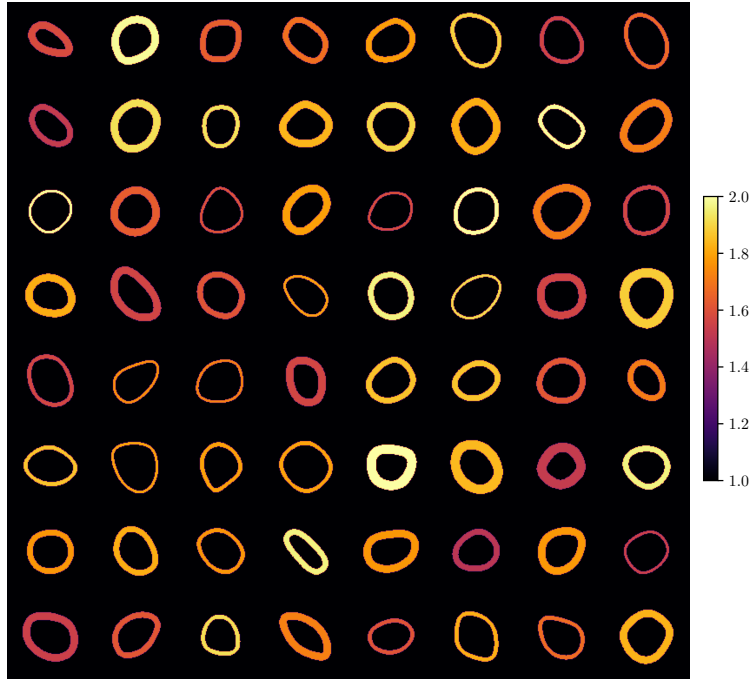}}
    }
    \end{adjustbox}
    \caption{Examples of the sound speed distributions based on idealized skulls used to train the learned optimizer. Each skull is created by summing up several circular harmonics of random amplitude and phase, and then assigned a random thickness between 2 and 10 pixels, and a random sound speed between 1.5 and 2 times the background value.}
    \label{fig:samples_from_dataset}
\end{figure}

The training set contains 9000 sound speed distributions while both the validation and test sets contain 1000 distributions. Several examples are shown in Fig. \ref{fig:samples_from_dataset}. As a physics-based loss function is used, the training data only contains sound speed distributions---no ground truth wavefields (e.g., generated using another PDE solver) are required. During training, the source distribution $\rho$ is always fixed to a single grid point with magnitude 10 at the bottom of the domain. 

From a practical point of view, it is not computationally feasible to perform backpropagation through a large number of iterative steps of the learned optimizer. To overcome this, the network is trained using a replay buffer and truncated backpropagation through time (TBPTT) \cite{jaeger2002tutorial}, where TBPTT is implemented by unrolling 10 iterations. The replay buffer is initially filled with 600 triplets $(c, u_k, h_k)$ containing sound speed distributions randomly selected from the training set. For each sound speed example, the wavefield, and hidden state are initialized to zero, while the iteration index $k$ is initialized as a random integer between 0 and the maximum number of iterations, in this case set to $T = 1000$.

During training, at each training step a mini-batch of $N$ triplets (containing examples with a range of different sound speed distributions and iteration indices) is randomly selected from the buffer. For each triplet, the loss is calculated over 10 iterative steps using Eq.~\eqref{eq:mse_physics_loss}, and the total loss is then summed over the mini-batch, where

\begin{equation}
    R = \frac{1}{N}\sum_{n} \sum_k L_{k}(u_{k,n},c_n,\rho) \enspace.
    \label{eq:linear_quadratic_tracking}
\end{equation}
Here $n\in \{0,\dots,N-1\}$ is the triplet index over the mini-batch, $ k \in \{k_n+1, \dots, k_n + 10\}$ is the iteration index, and $k_n$ is the starting iteration index for the $n$-th triplet. Finally a gradient descent step is performed over 10 unrolled iterations to update the network. 

The calculation of the loss $L_{k}$ using Eq.~\eqref{eq:linear_quadratic_tracking} is performed using a Fourier collocation spectral method  with a modified Laplacian that includes a PML. This allows the Sommerfeld radiation condition at infinity to be approximately satisfied while cropping the domain to a finite size $\Omega \subset \mathbb{R}^2$. The discrete formulation used is given in  Appendix \ref{sec:appendix}. 

After each training step, for each example in the mini-batch, one of the iterative steps is randomly selected, and a new triplet $(c, u_k, h_k)$ is stored back into the buffer replacing the previously selected triplet. For computational reasons, the residual is also stored alongside the triplet to avoid needing to recalculate this when the triplet is later re-drawn from the buffer. During training, if the iteration index of any of the triplets to be stored back into the buffer exceeds $T = 1000$, this is replaced with a new sound speed distribution randomly selected from the training set. The wavefield, hidden state, and iteration index for the new example are initialized to zero. The full training algorithm is given in Appendix \ref{app:training_algo}.

Note, the use of a small window for backpropagation can bias the network towards learning state representations which have short temporal dependency. While there are techniques for mitigating such bias \cite{tallec_unbiasing_2017}, we didn't find this to be a problem in practice. Storing the belief state $h_k$ in the replay buffer (rather than re-initializing it) has also been shown to improve the performance of recurrent networks trained using experience replay \cite{kapturowski_recurrent_2019}.

Gradient descent is performed using Adam with a batch size of 32, learning rate of $10^{-4}$, and gradient clipping at $1$. The biases of all convolutional layers are initialized to zero to minimize the risk of divergence of the wavefield in the early iterations. We also don't store triplets in the buffer if the loss goes above an arbitrary threshold value of $L=1$, as this suggests that the wavefield is diverging. We found this to be especially important in the early phase of training.

The network and training were implemented using PyTorch and parallelized using PyTorch Lightning \cite{falcon2019pytorch}. The training was performed using a cluster of 6 NVIDIA Tesla P40 graphics processing units (GPUs) in parallel. During training, at the end of every 10 epochs, the loss on the validation set was also evaluated, in this case by summing the loss at the last iteration $L_{T}$ over all examples in the validation set. However, in this case the source position $\rho$ for each example was moved to a random position on a circle to provide a simple test of network generalization. Since no input/output pairs were provided during training, inputs and outputs were scaled by a factor of $10^3$ and $10^{-3}$, respectively. These values were hand-tuned to roughly normalize the variance of the inputs and outputs across all iterations. Similarly, the loss function was amplified by a factor of $10^4$. The training was run for 1000 epochs (52k training steps), and the network with the lowest validation loss was selected. The total training time was approximately 21 hours. A summary of the network and training hyperparameters is given in Appendix \ref{app:hyperparameters}.
\section{Results}


\subsection{Model accuracy against a reference solution for the test set}

To evaluate the performance of the trained network, a series of tests were performed. First, the accuracy of the network for the (unseen) sound speed maps in the test set was evaluated by comparing the wavefields calculated after 1000 iterations of the learned optimizer against a reference solution. The reference solution was calculated using the open-source k-Wave acoustics toolbox \cite{treeby2010k}. To obtain a time-independent solution to the wave equation, the time-domain solver \verb=kspaceFirstOrder2DG= was used with a continuous wave sinusoidal source term, the solution was run to steady state, and the complex wavefield extracted by temporal Fourier transform. The wavefields were then normalized to an amplitude of 1 and phase of zero at the source location to account for the different source scaling and relative phase between the two models. The accuracy was computed using the relative $\ell_\infty$ and average RMSE error norms calculated as
\begin{equation}
    \ell_\infty = \frac{ \| u_\mathrm{predicted} - u_\mathrm{reference} \| _\infty}{\|  u_\mathrm{reference} \|_\infty} \enspace, \qquad \text{RMSE} = \sqrt{\frac{\| u_\mathrm{predicted} - u_\mathrm{reference} \|^2_2}{N}} \enspace,
\end{equation}
where $N$ is the total number of pixels in the wavefield. As the learned optimizer and k-Wave use different formulations for the PML, this region was excluded from the error calculations.

\begin{figure}
    \makebox[\textwidth][c]{\hspace*{-0.6cm}
        \scalebox{0.64}{\input{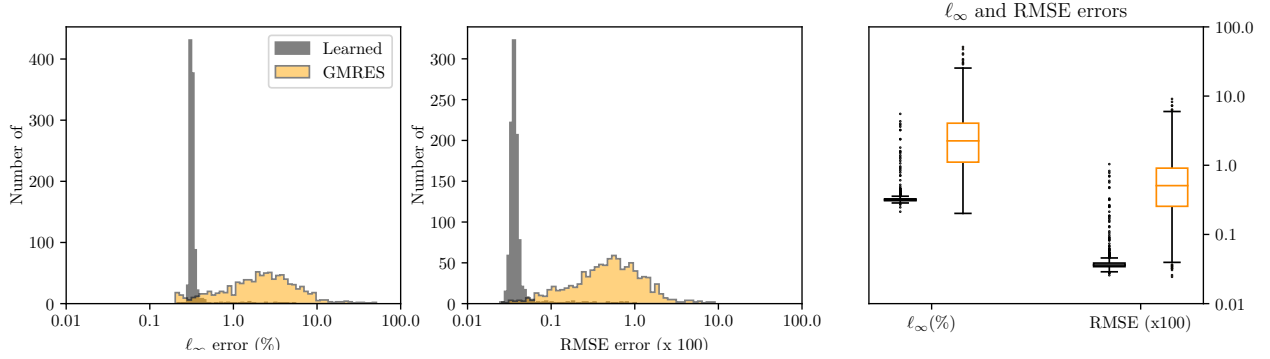}}
    }
    \caption{Errors in the wavefields predicted by the learned optimizer and the generalized minimal residual (GMRES) method shown as histograms (left and middle) and box plots (right). The errors are compared after 1000 iterations against a reference solution calculated using k-Wave for the 1000 sound speed distributions in the test set. The plotted RMSE errors are multiplied by 100 to move them to a similar scale. The learned optimizer is highly accurate, with a mean $\ell_\infty$ error of 0.36\%, and mean root mean square error (RMSE) of $4.6\times 10^{-4}$. }
    \label{fig:kwave_error_stats}
\end{figure}

\begin{figure}
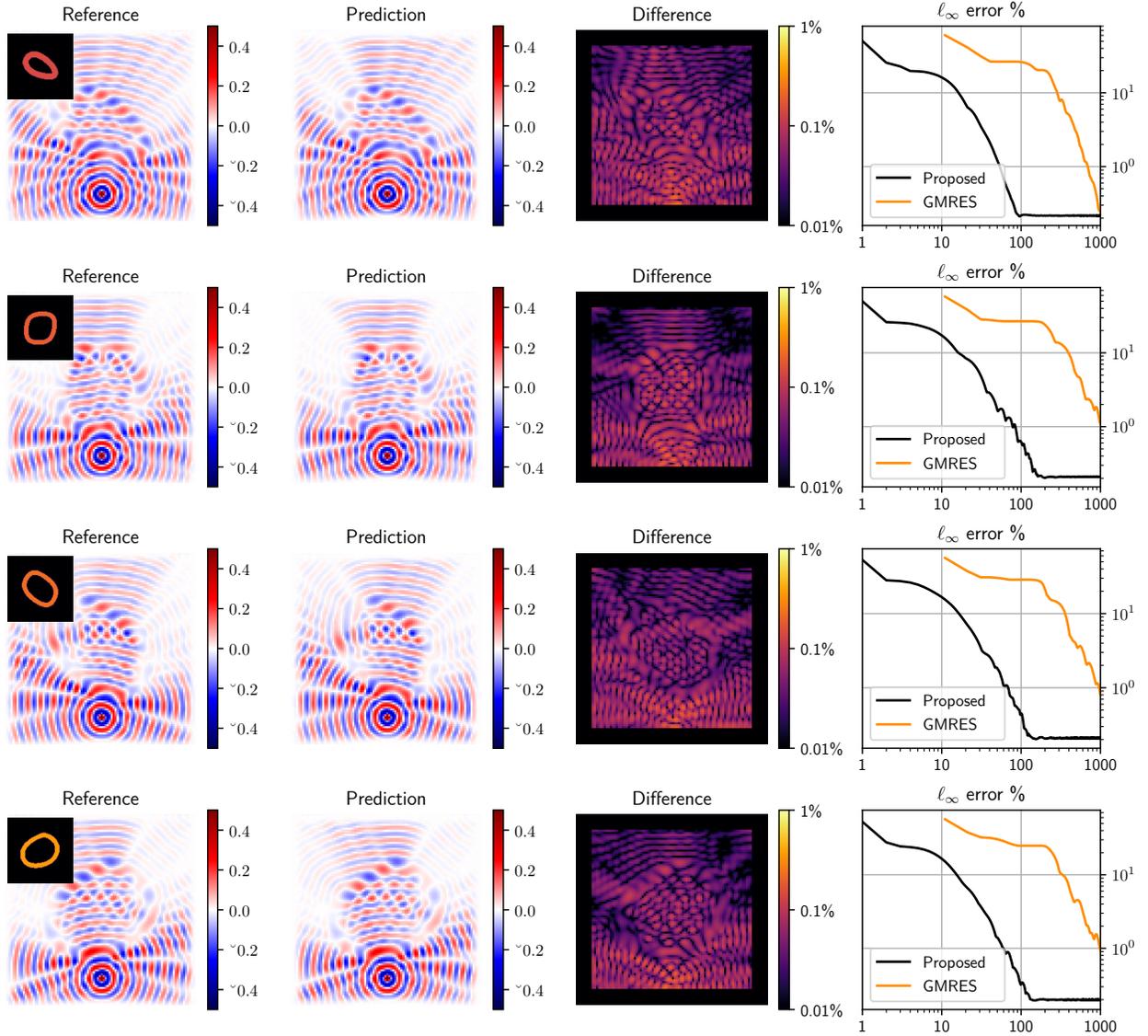

    \centering
    \makebox[\textwidth][c]{\hspace*{-1.2cm}
        \scalebox{0.67}{\input{images/examples/example_0.pgf}}
    }
    \makebox[\textwidth][c]{\hspace*{-1.2cm}
        \scalebox{0.67}{\input{images/examples/example_2.pgf}}
    }
    \makebox[\textwidth][c]{\hspace*{-1.2cm}
        \scalebox{0.67}{\input{images/examples/example_3.pgf}}
    }
    \makebox[\textwidth][c]{\hspace*{-1.2cm}
        \scalebox{0.67}{\input{images/examples/example_4.pgf}}
    }
    \caption{Four examples of simulations using idealized skull distributions randomly selected from the test set (insets shown top left). In each case, the reference solution is computed using k-Wave and shows very close agreement with the prediction using the learned optimizer (the real part of the wavefield is shown). For these examples, the $\ell_\infty$ error reaches a minimum within 200 iterations. \New{The convergence of GMRES is also shown for comparison: we reset the Krylov subspace every 10 iterations.}}
    \label{fig:test_samples}
\end{figure}

A histogram of the error norms for the 1000 examples in the test set is shown in Fig.\ \ref{fig:kwave_error_stats}, with four examples of the calculated wavefields given in Fig.\ \ref{fig:test_samples}. The predicted wavefields have very low errors compared to the reference solution, with a mean $\ell_\infty$ error of 0.36\%, and mean RMSE of $4.6\times 10^{-4}$. This demonstrates that the learned optimizer gives highly accurate results. Although not a focus of the current work, preliminary benchmarks show the learned model to be at least an order of magnitude faster than k-Wave for the same level of accuracy. 

\begin{figure}
    \centering
    \makebox[\linewidth]{
        \scalebox{0.68}{\input{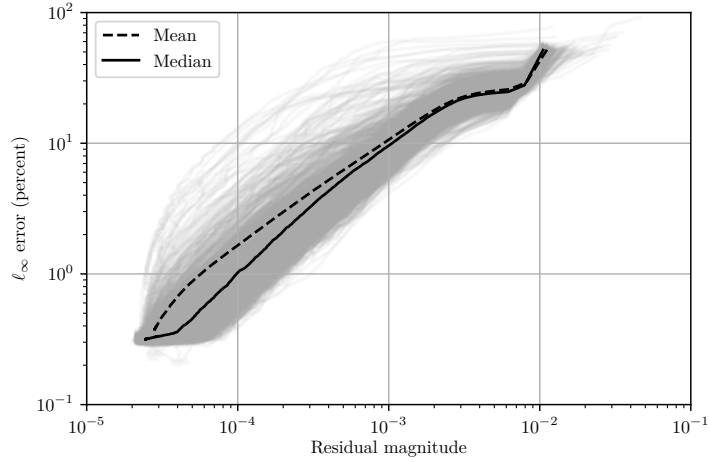}}
    }
    \caption{Trajectories showing the evolution of the $\ell_\infty$ error against the residual loss for the 1000 sound speed distributions in the test set. The $\ell_\infty$ error is calculated by comparison with k-Wave, and the residual is calculated using a physics loss term. The dashed and solid lines indicate the mean and the median of all traces, respectively.}
    \label{fig:error_vs_residual_traces}
\end{figure}

During the iterative procedure, it is possible to monitor the progression of the solution using metrics based on the computed residual. Figure \ref{fig:test_samples} gives four examples of the evolution of the residual with iteration number, and Fig.\ \ref{fig:evolution_of_residual_with_gmres} for all examples in the test set. Typically, a few hundred iterations are needed for the residual magnitude to reach a minimum. However, while a zero residual indicates convergence to the true solution, it is not immediately obvious how other values correspond to absolute accuracy. To investigate this, the evolution of the residual magnitude vs $\ell_\infty$ error with iteration number for the test set is plotted in Fig. \ref{fig:error_vs_residual_traces}. In general, the curves decrease with iteration number, meaning a lower residual magnitude gives a lower $\ell_\infty$ error for a given problem. However, while a high residual magnitude (e.g., $10^{-3}$) implies a high error, and a very low residual (e.g., $2\times 10^{-5}$) implies a low error, for intermediate values (e.g., $10^{-4}$), there is significant spread. This will be investigated further in future work.

In both Fig.\ \ref{fig:kwave_error_stats} (right panel) and Fig.\ \ref{fig:error_vs_residual_traces}, a small number of outliers can be seen with errors on the order of a few percent. In general, these outliers had a higher value for the final residual, and the evolution of the residual often displayed oscillations. Further investigation into these examples demonstrated the presence of a mode-like structure in the wavefield within and adjacent to the idealized skull, similar to whispering gallery modes. These examples were also much more difficult for k-Wave to compute, requiring at least twice as many time steps to reach an approximately steady state. Empirically it seems that examples in which highly resonant modes are supported take longer to converge, which suggests there may be a connection between the iterations and the time evolution of the field, as shown in Figs.\ \ref{fig:evolution_of_solution} and \ref{fig:real_skull_evolution}. The discrepancy between the mean and median error in the early steps of optimization as shown in Fig.\ \ref{fig:evolution_of_residual_with_gmres} also suggests that there are some particular sound speed distributions in the test set which are more challenging for the learned optimizer to solve. Interestingly, the same behavior was not observed for more complex sound speed distributions, e.g., based on real skulls.


\subsection{Comparison with GMRES\label{sec:GMRES}}

\begin{figure}
    \centering
    \makebox[\linewidth]{
        \scalebox{0.66}{\input{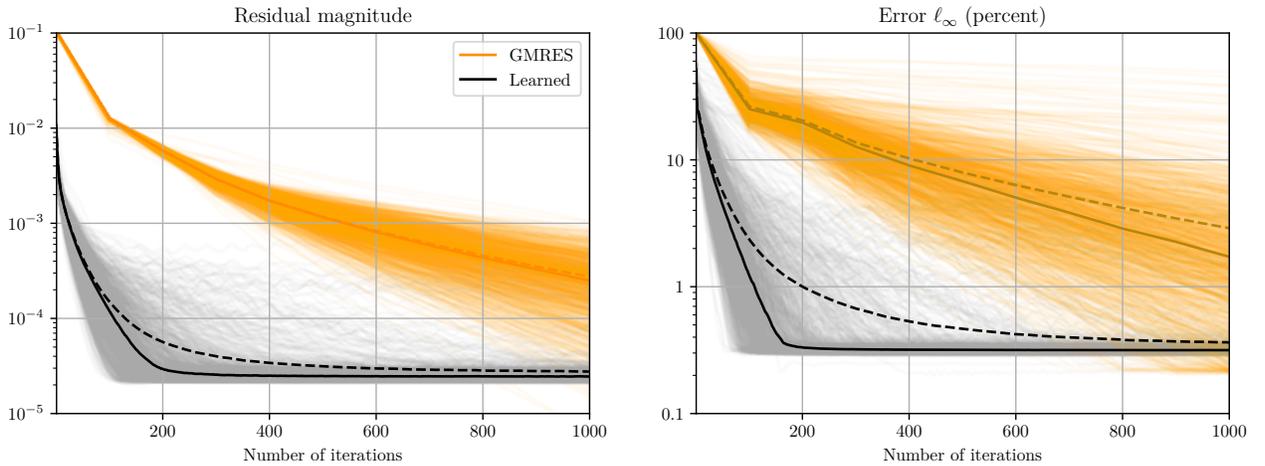}}
    }
    \caption{Progression of the optimization problem with the number of iterations for the learned optimizer (black) and the generalized minimal residual (GMRES) method (orange) for the 1000 sound speed distributions in the test set. The dashed and solid lines indicate the mean and the median of the individual traces, respectively. (left) Magnitude of the residual. (right) $\ell_\infty$ error compared to k-Wave. }
    \label{fig:evolution_of_residual_with_gmres}
\end{figure}

To benchmark against classical methods, we compared the learned iterative solver with the widely used GMRES method for the sound speed examples in the test set. Figure \ref{fig:kwave_error_stats} shows histograms and box plots of the $\ell_\infty$ and RMSE errors against k-Wave, while Fig.\ \ref{fig:evolution_of_residual_with_gmres} shows the progression of the residual and $\ell_\infty$ against iteration number. The learned optimizer outperforms the generic solver both in terms of convergence speed with iteration number, and in terms of accuracy for a given number of iterations. After 200 iterations, the learned optimizer reaches an average residual magnitude of about $5.6\times 10^{-5}$ and $\ell_\infty$ error of 1.0\%. In comparison, GMRES reaches an average residual magnitude of about $5.7 \times 10^{-3}$ and $\ell_\infty$ error of 20.4\%. Even after 1000 iterations (a five-fold increase), GMRES only reaches an average $\ell_\infty$ error of 2.8\%. 

\begin{figure}
    \centering
    \begin{subfigure}{\textwidth}
        \centering
        \makebox[\textwidth][c]{\hspace*{-0.9cm}
        \includegraphics[width=1.28\textwidth]{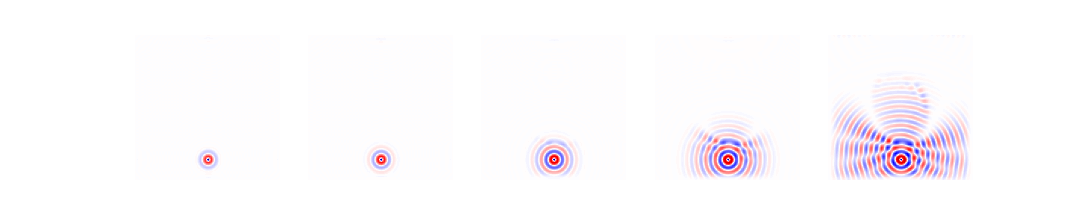}}
        \caption{Generalized minimal residual (GMRES) method.}
    \end{subfigure}
    \bigskip
    \begin{subfigure}{\textwidth}
        \centering
        \makebox[\textwidth][c]{\hspace*{-0.9cm}
        \includegraphics[width=1.28\textwidth]{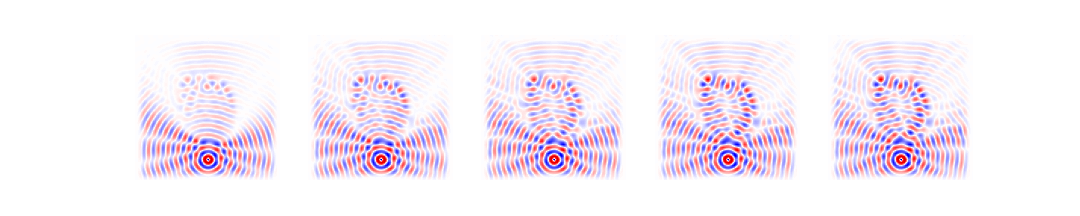}}
      \caption{Learned optimizer.}
    \end{subfigure} 
    \caption{Evolution of the solution after 10, 20, 50, 100, and 250 iterations for GMRES and the learned optimizer for a sound speed distribution from the test set (the real part of the wavefield is shown). The results for GMRES demonstrate the local nature of Krylov iterations based on the Helmholtz operator, which build the solution from the source location outwards. The results for the learned optimizer show a similar dynamic, but the updates cover the global domain much more rapidly.} 
    \label{fig:evolution_of_solution}
\end{figure}

The difference in the convergence rates can be understood by looking at the evolution of the solution with iteration number for a representative example as shown in Fig. \ref{fig:evolution_of_solution}. While both GMRES and the proposed solver construct the solution from the source location outwards, the spatial extent of the update made at each step by the two are very different: while GMRES tends to make very local updates (this is due to the nature of the Krylov iterations using a local forward operator, as discussed in \S\ref{sec:numerical_solution}), the learned iterative solver updates the solution over much larger spatial ranges.

\subsection{Network generalizability}
\label{sec:results_generalizability}

Having established the ability of the learned optimizer to give highly accurate solutions for the sound speed distributions in the test set, a preliminary investigation was performed into its generalization capabilities to solve previously unseen problems. Similar performance was also observed on a range of other examples analogous to the ones described below.

\begin{figure}
    \centering
    \makebox[\textwidth][c]{\hspace*{-1.2cm}
        \scalebox{0.67}{\input{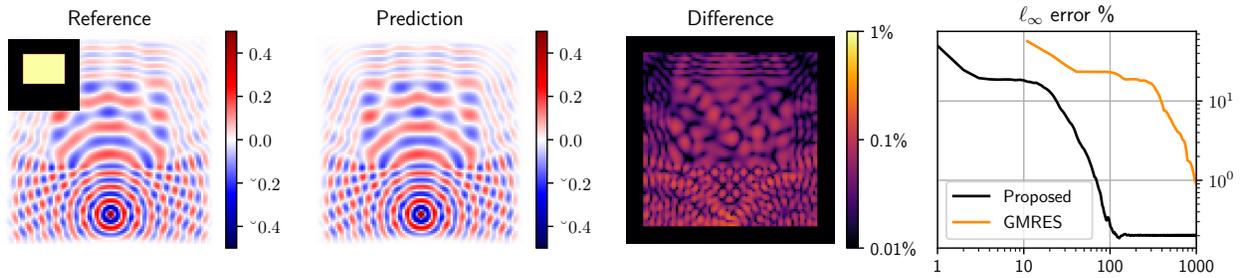}}
    }
    \caption{Simulation using a large rectangular heterogeneity (inset top left) which has a shape unlike the speed of sound distributions seen during training. The reference solution is computed using k-Wave and shows very close agreement with the prediction using the learned optimizer (the real part of the wavefield is shown). The relative $\ell_\infty$ error against the reference solution takes approximately 70 iterations to reach 1\% .}
    \label{fig:model_vs_kwave_rect_sample}
\end{figure}

First, we evaluated the model on a speed of sound distribution containing a rectangular region with a speed of sound of 2, to test the ability of the network to deal with large homogeneous speed of sound regions (recall during training that only idealized skull shapes were used). Figure \ref{fig:model_vs_kwave_rect_sample} shows the reference solution calculated using k-Wave, the prediction using the learned optimizer, and the evolution of the error with iteration number. For this example, the learned model reaches a very small final error on the order of 0.2\%, and reaches an error below 1\% extremely quickly (about 70 iterations). 

\begin{figure}
    \centering
    \makebox[\textwidth][c]{\hspace*{-1.1cm}
        \scalebox{0.67}{\input{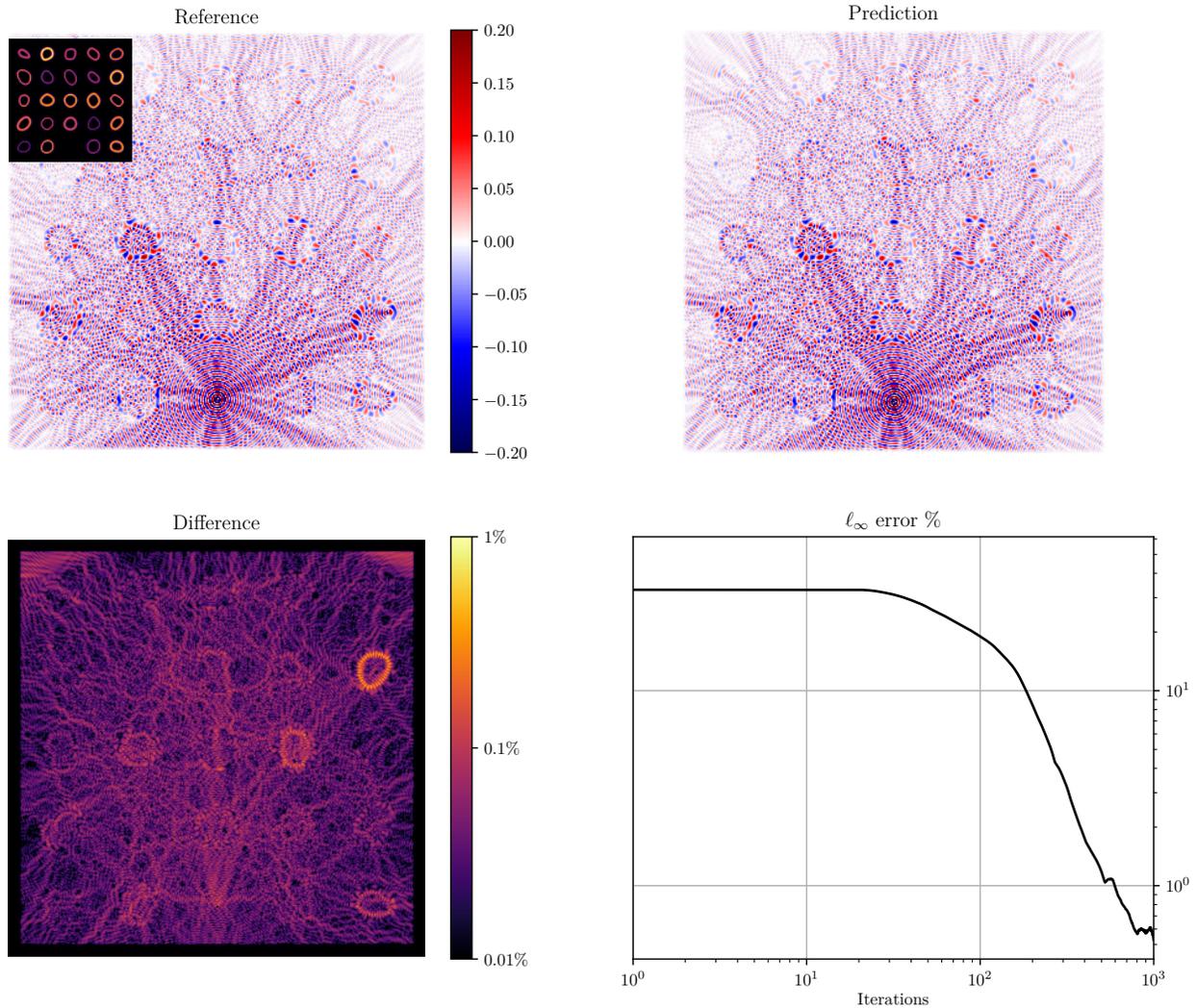}}
    }
    \caption{Test on a spatial domain much larger than that seen during training. The sound speed distribution is created by patching together 24 idealized skull distributions from the test set (inset top left). The reference solution is computed using k-Wave and shows very close agreement with the prediction using the learned optimizer, with errors well below 1\% (the real part of the wavefield is shown). The relative $\ell_\infty$ error against the reference solution takes approximately 600 iterations to reach 1\%. }
    \label{fig:patching}
\end{figure}

Second, we tested the ability of the network to generalize to larger domains. A large speed of sound distribution with 480 $\times$ 480 grid points was created by patching together 24 distributions from the test set. Figure \ref{fig:patching} shows the reference solution calculated using k-Wave, the prediction using the learned optimizer, and the evolution of the error with iteration number. Despite all the sound speed distributions in the training and validation sets having 96 $\times$ 96 grid points, the model is able to generalize to a much larger domain, reaching 1\% error within 600 steps. This suggests that the hard task of training on large 3D volumes containing whole skulls for clinical applications can possibly be entirely bypassed by learning the network weights using much simpler and smaller problems, for example, by training using small skull patches.

Finally, while the previous example suggests that the network is able to generalize to different domain sizes, it is unclear how much diversity in the training set is required to ensure that the network still converges to a satisfactory solution with an arbitrary sound speed or source distribution. To test this, we performed a representative simulation of the intracranial wavefield for a transcranial focused ultrasound stimulation experiment \cite{lee2016transcranial}. We used a large 512 $\times$ 512 speed of sound distribution generated from a transverse CT slice from an adult skull from the Qure.ai CQ500 Dataset \cite{chilamkurthy2018development} converted using the approach outlined in \cite{aubry2003experimental}. The source distribution was defined as a focused transducer represented by a 1D arc \cite{martin2016simulating} (recall that the network has only seen single-point sources in a fixed position during training). The transducer aperture diameter and radius of curvature were set to 60 mm, and the source frequency to 490 kHz. \New{We also recall that, due to the linearity of the Helmholtz equation, the wavefield resulting from an arbitrary source distribution can always be decomposed into the sum of wavefields produced by point sources. Therefore it is always possible to run the algorithm in parallel for a set of point-like source maps by decomposing the total source field, with the added benefit of decoupling the effect of each source on the total wavefield.}

\begin{figure}
    \centering
    \makebox[\textwidth][c]{\hspace*{-1.1cm}
        \scalebox{0.67}{\input{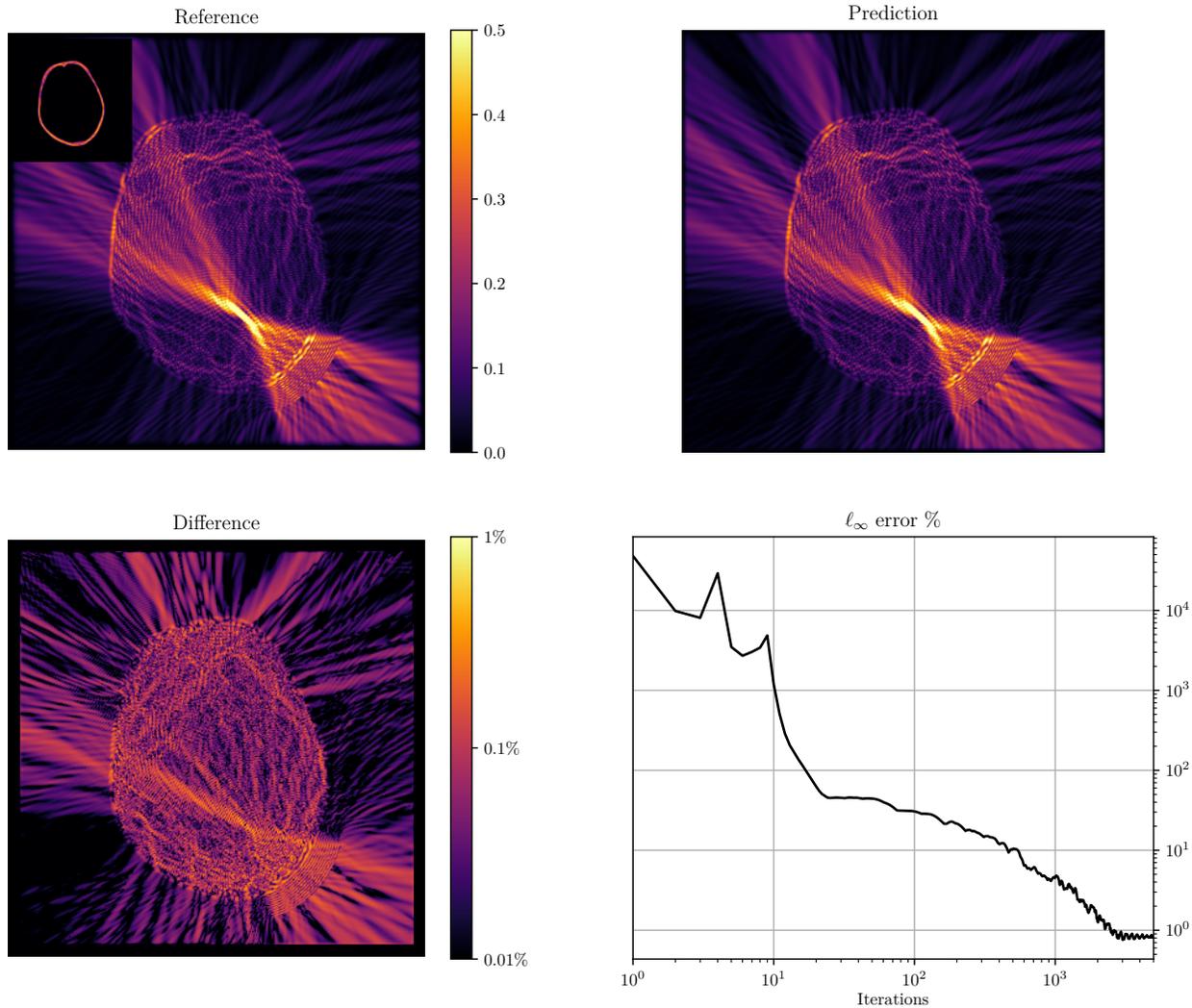}}
    }
    \caption{Simulation of a transcranial ultrasound field using a sound speed distribution generated from a clinical x-ray computed tomography (CT) slice (inset top left). The reference solution (top left) is computed using k-Wave and shows very close agreement with the prediction using the learned optimizer (top right), with errors below 1\% (bottom left). The absolute value of the complex wavefield is shown. The $\ell_\infty$ error takes approximately 3000 iterations of the optimizer to reach a minimum (bottom right).}
    \label{fig:real_skull_results}
\end{figure}

Figure \ref{fig:real_skull_results} shows the reference solution calculated using k-Wave, the prediction using the learned optimizer, and the evolution of the error with iteration number. Despite this example being well outside the training set (including a much larger spatial domain, a more complex speed of sound distribution, and a distributed source in an unseen position), the relative $\ell_\infty$ error compared to k-Wave is very low at 0.8\%. This shows that the trained model can be used to solve problems of the scale and complexity needed to make the clinical problem tractable, albeit currently in 2D.

\begin{figure}
    \centering
    \includegraphics[width=\textwidth]{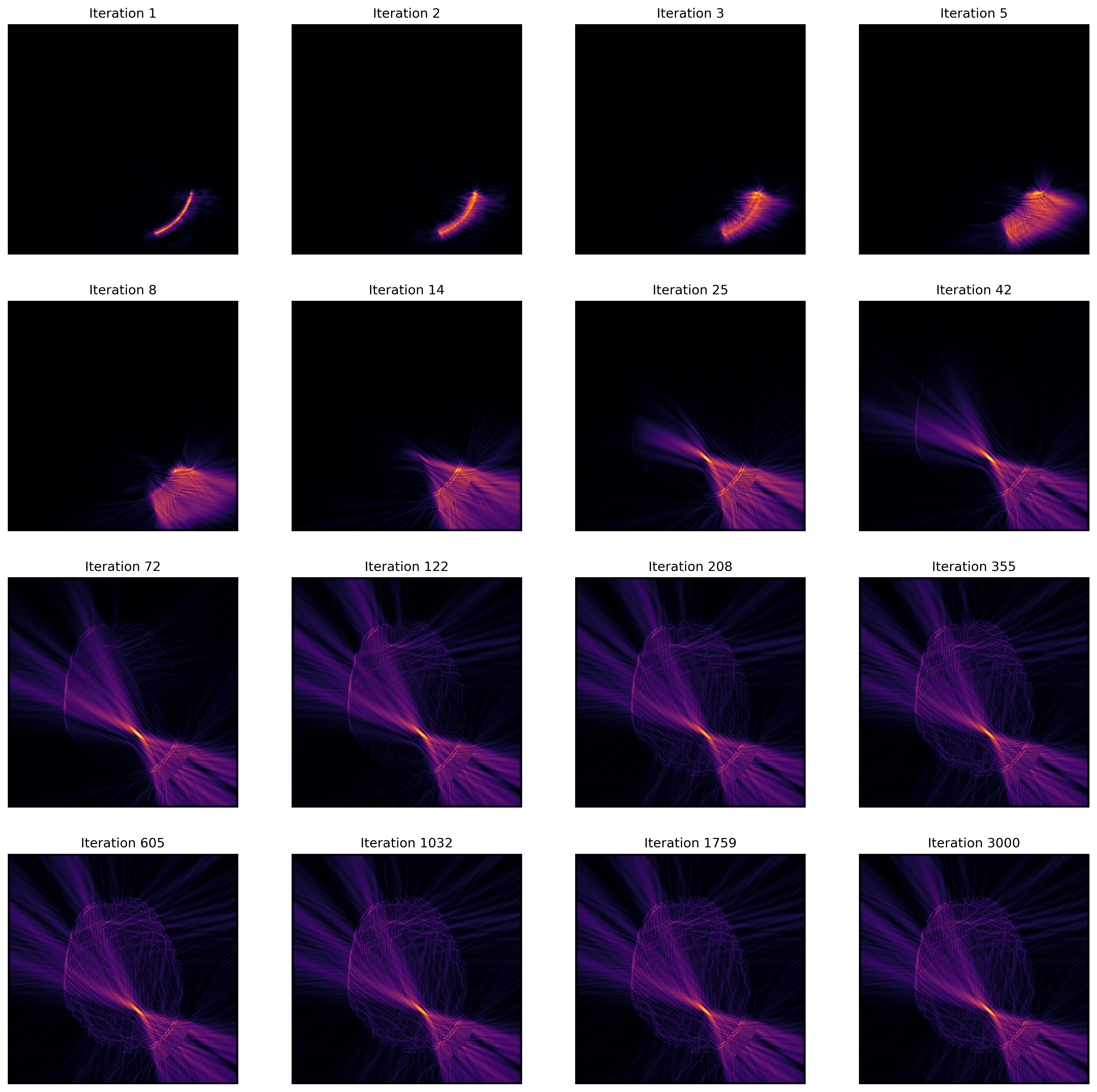}
    \caption{Evolution of the wavefield predicted by the learned optimizer for different iteration numbers for the transcranial ultrasound example given in Fig.~\ref{fig:real_skull_results}. The position of the focus and the focal pressure converge with 20 iterations, and most of the complex structure in the field can be seen after just 200 iterations. Note, the colormap of each subpanel is normalized by its own maximum.}
    \label{fig:real_skull_evolution}
\end{figure}

The evolution of the solution shown in Fig.\ \ref{fig:real_skull_results} with iteration number is shown in Fig.~\ref{fig:real_skull_evolution}. The solution is constructed from the arc-shaped source outwards. While it takes approximately three thousand iterations for the $\ell_\infty$ error across the whole domain to reach a minimum, most of the complex structure in the field can be seen after just 200 iterations. Moreover, the position of the focus and the focal pressure (which are not strongly affected by reverberation within the skull for this example) converge extremely quickly, within just 20 iterations. For the purpose of real-time treatment planning (for example, where several candidate positions for the ultrasound transducer may be being evaluated), the ability to generate a reasonable approximation for the acoustic field around the focus in a very short time, which can then be improved by letting the model continue to iterate, is highly desirable. (See Supplementary Material for an example.)

\section{Summary and discussion}

A lightweight neural network based on a modified UNet is proposed as a fully learned iterative optimizer and used to solve the heterogeneous Helmholtz equation. The network architecture is motivated by multi-scale solvers which utilize multi-scale memory, and Markov decision processes that utilize a belief state to make them fully observable. The network is trained using a physics-based loss function: no explicit knowledge of the true solution is given beyond the implicit consistency with the known source and sound speed.

The training set consists of idealized skulls in a small domain with 96 $\times$ 96 grid points, which makes training time manageable (less than one day using 6 GPUs). The learned optimizer shows excellent performance on the test set, and is capable of generalization well outside the training set, including to much larger domains, more complex sound speed distributions, and more complex source distributions.

In the current work, the network is used to solve the lossless Helmholtz equation in 2D. In future, we will look to extend this to 3D, and to include the effects of changes in mass density, acoustic absorption, and potentially the effects of acoustic nonlinearity. The examples given in \S3.3 suggest that the model weights could be trained on small 3D patches, which will be essential to make the training tractable.

\New{The neural network has not been constrained in any way and therefore it is hard to state any \textit{a priori} guarantees on its convergence. In general, the convergence of repeatedly applying neural networks with weight-tying is still an open problem (see \cite{kawaguchi2021simple}). However, the network could be constructed in such a way to ensure at least convergence, for example, by taking optimal step-sizes.} \New{Regarding convergence speed, standard iterative solvers often rely on some form of preconditioning to achieve convergence in a small number of iterations. Preconditioners could also be used to enhance the performance of learned solvers, either by applying them at the residual calculation stage or by using them to improve the spectral properties of the physics loss.}

One challenge in the formulation that may need addressing is the limited number of unrolling steps used in the training phase. While a replay buffer has proven useful to extend the training horizon well beyond this limit, it may still induce biases or unwanted phenomena such as state representation drift \cite{kapturowski_recurrent_2019}. This could be mitigated by borrowing ideas from the reinforcement learning community, which often deals with large or even infinite horizon tasks. In particular, Q-learning methods \cite{sutton_reinforcement_1998}, such as temporal difference learning, can theoretically take into account the entire sequence of possible future states when providing the loss for a given output, by estimating the future loss. 
Furthermore, as they are designed to work with externally given reward signals, the loss function on which the network is trained doesn't need to be differentiable. This makes it possible to use more elaborate training strategies, such as imposing a monotonically decreasing residual norm during inference. 

Since we are mostly interested in the solution in a restricted region of space, such as the brain for transcranial applications, an interesting extension of the method would be to include a spatial estimate of the uncertainty of the solution at each step \cite{zhu_physics-constrained_2019}. While being useful in and of itself, this may also allow the design of training procedures where the network focuses on solving the problem in the spatial region of interest.

In our experiments, the complexity of the network was kept very low by using a small number of parameters, and this is partially responsible for the good generalization performance. On the \New{one} hand, having a small number of parameters reduces the capacity of the network and as a consequence the speed at which a problem is solved.  Conversely, having a large number of parameters may overspecialize the network on the distribution of problems encountered in training, which may require a more diverse training dataset to restore its generalization capabilities. Furthermore, the validation loss exhibits large oscillations throughout training, making evaluation and checkpointing of the network parameters subject to a low validation loss a crucial part of the training process. \New{Lastly, the neural network was trained with a fixed frequency relative to the grid spacing (i.e., a fixed number of points per wavelength in the background medium). Therefore it is applicable to all problems as long as the spatial discretization is chosen to satisfy this constraint. For arbitrary frequencies (or, equivalently, grid resolutions) the network currently needs to be retrained or fine-tuned. However this requirement may be relaxed in the future by training the solver on a suitable range of frequencies. While this may reduce the performance of the resulting network, it could also act to regularize the network and allow training larger networks.}
As part of future work, we will aim to understand and disentangle the effect of dataset diversity, network complexity, \New{training strategies}, and the generalization properties of the network.

Lastly, although the neural network solver already provides an advantage over many standard numerical methods by being fully differentiable, the computational performance of the network still needs to be properly assessed. In particular, profiling and optimization of the deployed network, and a comparison to more widespread solving procedures including a measure of their scaling properties for different problem sizes are required. While there may be several small design ideas that can be used to reduce the computational complexity of the network, such as learning the Laplacian operator \cite{long2018pde} rather than evaluating it using spectral methods, a fair comparison will require a careful implementation of the method.

\section*{Acknowledgments}
Funding: This work was supported by the Engineering and Physical Sciences Research Council (EPSRC), UK. This work was also supported by European Union’s Horizon 2020 Research and Innovation Program H2020 ICT 2016-2017 (as an initiative of the Photonics Public Private Partnership) under Grant 732411.

\small{
    \bibliography{references}
    \bibliographystyle{ieeetr}
}
\appendix
\newpage
\section{Discrete solution of the forward operator}
\label{sec:appendix}

The calculation of the residual in Eq.\ \eqref{eq:iterative_solver} and the loss function in Eq.\ \eqref{eq:linear_quadratic_tracking} require the discretization of the heterogeneous Helmholtz equation in Eq.\ \eqref{eq:helmholtz}. To allow the Sommerfeld radiation condition at infinity to be approximately satisfied while cropping the domain to a finite size $\Omega \subset \mathbb{R}^2$, a perfectly matched layer or PML is used as defined in \cite{bermudez_optimal_2007}. Assuming the original bounded domain is rectangular with size $2L_x$ by $2L_y$, the domain is extended to a size of $2(L_x+\Delta L)$ by $2(L_y+\Delta L)$, where $\Delta L$ is the thickness of the PML on each side of the domain (see Fig.\ \ref{fig:domain_definition}). The derivative operators are then transformed to introduce absorption within the extended part of the domain:
\begin{equation}
    \frac{\partial}{\partial \eta} \to \frac{1}{\gamma_\eta}\frac{\partial}{\partial \eta} \enspace,
    \label{eq:deriv_with_pml}
\end{equation}
where $\eta = x,y$ and
\begin{equation}
    \gamma_\eta = \left\{ 
        \begin{aligned}
            1, & \quad \textnormal{ if } |\eta|<L_\eta \\
            1 + \frac{j}{\omega} \sigma(x), & \quad \textnormal{ if }  L_\eta \leq |\eta|<L_\eta + \Delta L \enspace.
        \end{aligned}
        \right.
\end{equation}
The absorption profile $\sigma$ grows quadratically within the perfectly matched layer according to \cite{bermudez_optimal_2007}
\begin{equation}
    \sigma(\eta) = \sigma_{\text{max}}\left(1 - \frac{\eta}{L} \right)^2.
\end{equation}
The Laplacian including the PML then becomes 
\begin{equation}
    \hat \nabla^2 = \frac{1}{\gamma_x}\frac{\partial}{\partial x}\left( \frac{1}{\gamma_x}\frac{\partial}{\partial x} \right) +  \frac{1}{\gamma_y}\frac{\partial}{\partial y}\left( \frac{1}{\gamma_y}\frac{\partial}{\partial y} \right) 
    = \sum_{\eta = x,y} \left[  \frac{1}{\gamma_\eta^2} \frac{\partial^2}{\partial\eta^2}- \frac{\gamma_\eta'}{\gamma_\eta^3}\frac{\partial}{\partial \eta} \right] \enspace,
    \label{eq:modified_operators_with_pml}
\end{equation} 
where $\gamma_\eta' = \frac{\partial}{\partial \eta}\gamma_\eta$, which is computed analytically. To discretize Eq.\ \eqref{eq:modified_operators_with_pml}, the Fourier collocation spectral method is used \cite{boyd2001chebyshev}, where
\begin{equation}
    \frac{d}{d\eta}f(\eta) = \mathcal{F}_\eta^{-1} \left\{  ik_\eta \mathcal{F}_\eta \left\{ f(\eta) \right\} \right\} \enspace.
\end{equation}
Here $k_\eta$ are the wavenumbers in the $\eta$-direction (see e.g., \cite{boyd2001chebyshev}), and $\mathcal{F}$ and $\mathcal{F}^{-1}$ represent the forward and inverse Fourier transform, in this case computed using the fast Fourier transform (FFT). Finally, this gives

\begin{equation}
    \hat \nabla^2 f(x,y) =  
    \sum_{\eta = x,y}   \left[\frac{1}{\gamma_\eta^2} \mathcal{F}_\eta^{-1} \left\{ (-k_\eta^2) \mathcal{F}_\eta \left\{ f(x,y) \right\} \right\} -\frac{\gamma_\eta'}{\gamma_\eta^3} \mathcal{F}_\eta^{-1} \left\{ (ik_\eta) \mathcal{F}_\eta \left\{ f(x,y) \right\}\right\}\right]\, \enspace.
\end{equation} 
For a numerical calculations presented in the current work, the PML size $\Delta L$ is set to 8, and the absorption parameter $\sigma_\mathrm{max}$ is set to 2.


\newpage
\section{Training algorithm}
\label{app:training_algo}
The algorithm used for training is given in Algorithm. \ref{algo:training}

\begin{algorithm}[bh!]
\DontPrintSemicolon
  \KwData{Train set $\mathcal{C}$, Validation set $\mathcal{V}$, source $\rho$, buffer size $N_s$, maximum iteration $T$, unrolling steps $t$, batch size $N_b$}
  \tcc{Initialize replay buffer}

  \For{ $n \in [1,\dots,N_s]$}
   {
	$k \sim \mathcal{U}(0,T-1)$\;
	$c \sim \mathcal{C}$\; 
	Initialize $u_k,h_k$ with zeros\;
   	Store $(c,u_k,h_k)$ in the buffer $B$\;
   }

  \tcc{Training loop}
  $\hat V \leftarrow \infty$ \tcp*{Initialize best validation loss to infinity}
  \For{epoch $\in \{1,\dots,N_{epochs}\}$}
  {	
	\tcc{Train network for a single epoch}
	\For{step $ \in \{1,\dots,N_{steps}\}$}
	{
		Sample a batch of $N_b$ triplets $(c,u_k,h_k)$ from $B$\;

		\tcc{For each triplet}
		\For{$n \in \{1,\dots,N_b\}$}
		{
			\tcc{Unroll $t$ steps}
			\For{$i \in \{k+1,\dots,k+t\}$} 
			{
				$e_{i-1} \leftarrow A(c)u_{i-1} - \rho$\;
				$u_i \leftarrow u_{i-1} + f_\theta(u_{i-1},e_{i-1},h_{i-1})$\;
			}
			$L_n = \sum_i L_{i,n}$ \tcp*{Estimate the loss over $t$ steps}\;

			\tcc{Update replay buffer}
			$i \sim \mathcal{U}(k+1,k+t)$  \tcp*{Sample one of the $t$ iterations}
			\If{$i < T$}
			{
				Store $(c,u_i,h_i)$ in the buffer $B$ in place of its original triplet\;
			}
		    	\Else
		    	{
				$c \sim \mathcal{C}$\; 
				Initialize $u_0,h_0$ with zeros\;
			   	Store $(c,u_0,h_0)$ in the buffer $B$ in place of its original triplet\;
		    	}
		}
		$R  \leftarrow \frac{1}{N_b} \sum_n L_n$  \tcp*{Estimate the loss over $t$ steps}
		$\theta \leftarrow SGD(\theta, \nabla_\theta R) $   \tcp*{Update parameters via gradient descent}
	}
	\tcc{Evaluate the network on the validation set every 10 epochs}
	\If{$mod(epoch,10) = 0$} 
	{	
		$L_{val} \leftarrow 0$\;
		\For{every $c \in V$}
		{
			Initialize $u_0,h_0$ with zeros\;
			Initialize the source map $\rho$ with a point source at a random position\;
			\For{$i \in \{1,\dots,T\}$} 
			{
				$e_{i-1} \leftarrow A(c)u_{i-1} - \rho$\;
				$u_i  \leftarrow u_{i-1} + f_\theta(u_{i-1},e_{i-1},h_{i-1})$\;
			}
			$L \leftarrow L_{T,n}$ \tcp*{Estimate the loss at the last iteration}
			$L_{val} \leftarrow L_{val} + L$ \tcp*{Add to the total loss over the validation set}

		}
		\If{$L_{val} < \hat V$}
		{
			$\hat V \leftarrow L_{val}$\;
			Save the current network parameters $\theta$\;
		} 
	}
  }
\caption{Training procedure}
\label{algo:training}
\end{algorithm}


\newpage
\section{Model and training hyperparameters}
\label{app:hyperparameters}

A summary of the hyperparameters used in the network and for training is given in Tables \ref{tab:model_hyperparameters} and \ref{tab:training_hyperparameters}. \New{All the convolutional layers were initialized using the Xavier method with gain 0.02 \cite{glorot2010understanding}.}

\begin{table}[bh!]
\centering
\caption{Model hyperparameters}
\begin{tabular}[t]{lc}
\hline
\textbf{Parameter}  & \textbf{Value} \\ 
\hline
activation function & PReLU          \\
depth               & 4              \\
channels per layer  & 8              \\
channels per state  & 2              \\ 
\hline
\end{tabular}
\label{tab:model_hyperparameters}
\end{table}%

\begin{table}[bh!]
\centering
\caption{Training hyperparameters}
\begin{tabular}[t]{lc}
\hline
\textbf{Parameter}    & \textbf{Value} \\ 
\hline
batch size            & 32             \\
buffer size           & 600            \\
gradient clipping     & 1              \\
learning rate         & $10^{-4}$      \\
weight decay          & $10^{-6}$      \\
window size for TBPTT & 10             \\ 
\hline
\end{tabular}
\label{tab:training_hyperparameters}
\end{table}%

\end{document}